\documentclass[reprint,preprintnumbers,footinbib,superscriptaddress,amsmath,amssymb,floatfix,prl]{revtex4-1}

\usepackage{graphicx,color}           

\def\mT{{\ {\rm mT}}}                       

\def\Hz{{\ {\rm Hz}}}                       
\def\GHz{{\ {\rm GHz}}}                     

\def\nK{{\ {\rm nK}}}                       

\def\El{E_L}                            
\def\kl{k_L}                            
\def\Rb87{^{87}\text{Rb}}                     
\def\Na23{^{23}\text{Na}}                     
\def\Li6{^{6}\text{Li}}                       

\def\ket#1{\mathinner{|{#1}\rangle}}

{\catcode`\|=\active
  \gdef\Braket#1{\left<\mathcode`\|"8000\let|\BraVert {#1}\right>}}
\def\BraVert{\egroup\,\mid@vertical\,\bgroup}

\begin{document}

\title{Realistic Rashba and Dressehaus spin-orbit coupling for neutral atoms}

\author{D.~L.~Campbell}
\affiliation{Joint Quantum Institute, National Institute of Standards and Technology, and University of Maryland, Gaithersburg, Maryland, 20899, USA}

\author{G.~Juzeli\={u}nas}
\affiliation{Institute of Theoretical Physics and Astronomy, Vilnius University, A. Go\v{s}tauto 12, Vilnius LT-01108, Lithuania}

\author{I.~B.~Spielman}
\affiliation{Joint Quantum Institute, National Institute of Standards and Technology, and University of Maryland, Gaithersburg, Maryland, 20899, USA}

\date{\today}

\begin{abstract}
We describe a new class of atom-laser coupling schemes which lead to spin-orbit coupled Hamiltonians for ultra-cold neutral atoms.  By properly setting the optical phases, a pair of degenerate pseudospin (a linear combination of  internal atomic) states emerge as the lowest energy eigenstates in the spectrum, and are thus immune to collisionally induced decay.  These schemes use $N$ cyclically coupled ground or metastable internal states.  We specialize to two situations: a three and four level case where the latter adds a controllable Dresselhaus contribution.  We describe an implementation of the four level scheme for $\Rb87$ and analyze its sensitivity to typical laboratory noise sources.  Lastly, we argue that the Rashba Hamiltonian applies only in the large intensity limit since any laser coupling scheme will produce terms non-linear in momentum that decline with intensity.
\end{abstract}

\maketitle

%
%
Spin-orbit (SO) coupling is essential for realizing topological insulators, non-interacting fermionic systems with topological order~\cite{Kane2005,Hasan2010},  and yet in other contexts it leads to parasitic effects such as reduced spin coherence times~\cite{Koralek2009}.  As with the progression from the single-particle integer quantum Hall effect (IQHE) to the interaction driven fractional quantum Hall effects (FQHEs), the next important step is realizing the strongly interacting cousins to the topological insulators, of which topological superconductors are a first example~\cite{Schnyder2008,Sau2010}.  Since ultracold atoms lack intrinsic SO coupling, numerous techniques for generating SO coupling (generally equivalent to nonabelian gauge potentials~\cite{Dalibard2010}) with optical~\cite{Osterloh2005,Ruseckas2005,Zhu2006,Liu2009,Juzeliunas2010} and now rf~\cite{Goldman2010a} fields have been suggested, one of which was recently implemented~\cite{Lin2011}.

Current proposals for realizing SO coupling suffer from two primary limitations.  First, the pair of dressed spin states comprising the effective spin-$1/2$ system are not the two lowest energy states, so collisional deexcitation~\cite{Spielman2006} can rapidly transfer population into the ground state~\cite{Williams2011}.  Second, the required tripod coupling scheme is difficult to directly realize in alkali atoms~\footnote{Reimplementing the tripod or tetrapod schemes all within the electronic ground state is also possible~\cite{Juzeliunas2010,Dalibard2010}.}.  In this Rapid Communication, we introduce a class of laser coupling techniques that overcome these difficulties, and we explore the departure of such models from the ideal case (suitable only at infinite laser intensity).

In condensed matter systems, SO coupling links the linear or crystal (not orbital) momentum $\hbar \bf k$ to the spin of an electron, for example.  For systems confined to two dimensions (2D), terms linear in momentum can be represented as a sum of Rashba $\alpha\left(\check\sigma_x k_y - \check\sigma_y k_x\right)$ and Dresselhaus $\beta\left(\check\sigma_x k_y + \check\sigma_y k_x\right)$ SO couplings~\cite{Dresselhaus1955,Bychkov1984},  where $\check\sigma_{x,y,z}$ are the Pauli matrices.  Proposals for creating SO coupling with neutral atoms use lasers to link states of different momentum and spin.  Because these lasers impinge from discrete directions, the system lacks the continuous rotational symmetry of the pure Rashba Hamiltonian anticipated in earlier works~\cite{Stanescu2008,Liu2009,Juzeliunas2010}.  We show how a perturbative treatment restores the system's N-fold rotational symmetry and demonstrate that these pure couplings are only exact in limit of infinite laser-atom coupling strength. 

\begin{figure}[b!]
\begin{center}
\includegraphics[width=3.3in]{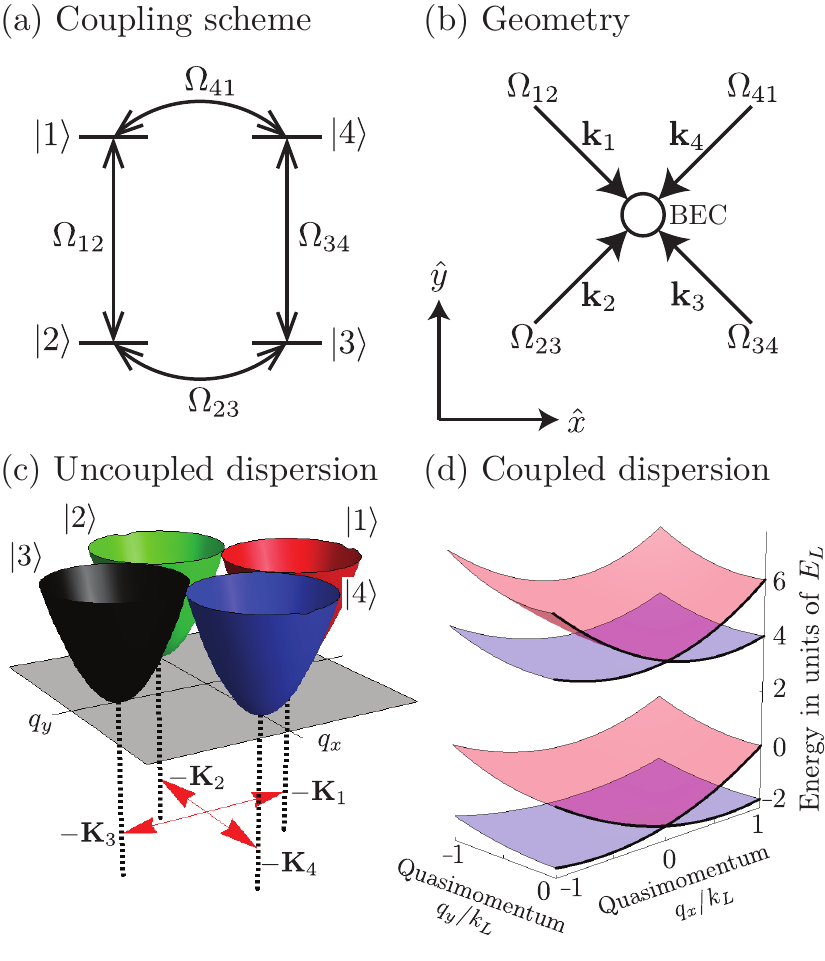}
\end{center}
\caption[Geometry and level diagram]{Four level scheme. (a) Effective coupling between four internal ground or metastable atomic levels.  (b) Spatial orientation of coupling fields. (c) Uncoupled eigenenergies for $\Omega=\epsilon=0$.  The four free parabolae are displaced by ${\bf K}_j$ in the $q_x$-$q_y$ plane. (d) Dispersion of four {\it dressed} states for $\Omega=3 E_R$, $\epsilon=0$, and $\bar\gamma=\pi/4$, showing the expected Dirac points, one for each pair of dressed bands.}
\label{fig_setup}
\end{figure}

%
%
We consider $N$ ground or metastable atomic ``spin'' states $\left\{\ket{1},\ket{2},\cdots,\ket{N}\right\}$ cyclicly coupled together with complex valued matrix elements $\Omega_{j+1,j}=-\Omega\exp\left[i \left({\bf k}_j\!\cdot\!{\bf x} + \gamma_j\right)\right]$ representing plane waves that link consecutive states $\ket{j}$ to $\ket{j+1}$.  Here, $\Omega$ describes the optical coupling strength; $\hbar{\bf k}_j$ and $\gamma_j$ are the respective discreet momentum and phase acquired in the $j\rightarrow j+1$ transition.  Throughout this manuscript we apply ``periodic boundary conditions'' $\ket{N+1}=\ket{1}$ for spin states.   See, for example, the 4-state topology in Figure~\ref{fig_setup}a.

Including the motional degrees of freedom, the many-body Hamiltonian
\begin{align}
\hat {\mathcal H} =& \int\frac{d^2{\bf k}}{(2\pi)^2}\sum_{j=1}^{N}\Bigg\{\left[\frac{\hbar^2\left|{\bf k}\right|^2}{2 m} + (-1)^j\frac{\epsilon}{2}\right]\hat\phi_j^\dagger({\bf k})\hat\phi_j({\bf k}) \nonumber\\
&-\frac{\Omega}{2}\left[e^{i\gamma_j}\hat\phi_{j+1}^\dagger({\bf k}+{\bf k}_j)\hat\phi_j({\bf k}) + \textrm{h.c.}\right]\Bigg\}\label{eq:main}
\end{align}
describes a system of atoms with mass $m$ in 2D absent the ubiquitous confining potential.  Here, $\{\phi^\dagger_j({\bf k})\}$ is the spinor field operator describing the creation of an atom with momentum ${\hbar \bf k}$ in internal state $\ket{j}$; for even $N$, we introduce $\epsilon$, describing a detuning of alternating sign.

In what follows, we require that $\sum {\bf k}_i=0$, so that no momentum is transferred to an atom during a $\ket{1}\rightarrow\cdots\rightarrow\ket{N}\rightarrow\ket{1}$ transition.  We define the momenta-exchange with differences ${\bf k}_j={\bf K}_{j+1}-{\bf K}_j$ and require $\left\{{\bf K}_j\right\}$ to have zero average.  Moreover, $\bar\gamma$ may replace the phase $\gamma_j$ of each state vector $\ket{j}$, where $\bar\gamma = \sum_i\gamma_i/N$ \footnote{The displacement vectors ${\bf K}_j = \sum_l l{\bf k}_{l+j-1}/N$ and phases $\theta_j=\sum_{l=1}^{j-1}(\gamma_l-\bar\gamma)$ define these transformations explicitly.}, without loss of generality.

%

\begin{figure}[t!]
\begin{center}
\includegraphics[width=3.5in]{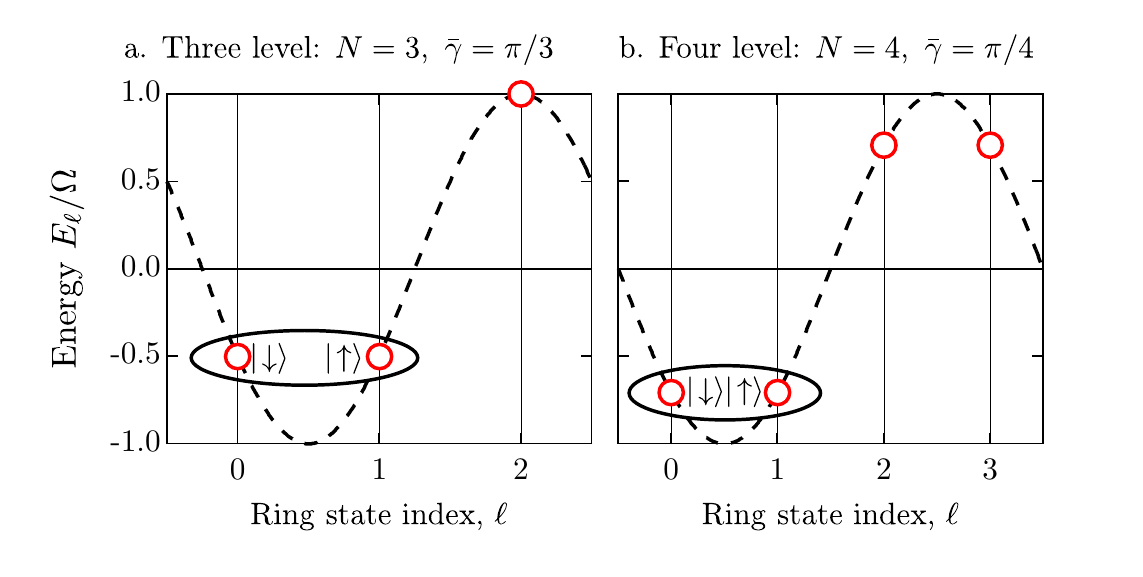}
\end{center}
\caption{Eigenenergies $E_\ell$ of the coupling term in Eq.~(\ref{EqnDisplaced}) showing the cosinusoidal energies evaluated at integer $\ell$.  (a) Three level case with $\bar\gamma=\pi/3$ and (b) Four level case with  $\bar\gamma=\pi/4$.  For both cases we identify pseudospin states $\ket{\uparrow}$ and $\ket{\downarrow}$ with the lowest energy pair of states.}
\label{fig_energies}
\end{figure}

With the substitution $\hat{\bar\varphi}^\dagger_j({\bf q}) = \hat\phi^\dagger_j({\bf q}+{\bf K}_j)$, the Hamiltonian [Eq.~(\ref{eq:main})] separates into an integral $\int \sum_{j,j^\prime}\hat{\bar\varphi}^\dagger_j({\bf q}){\bar H}_{j,j^\prime}({\bf q}) \bar\varphi_{j^\prime}({\bf q})d^2{\bf q}/(2\pi)^2$ over $N\!\times\!N$ blocks
\begin{align}
\bar H_{j,j^\prime}({\bf q}) =& \frac{\hbar^2\left|{\bf q}+{\bf K}_j
\right|^2}{2 m}\delta_{j,j^\prime} + (-1)^j\frac{\epsilon}{2}\delta_{j,j^\prime}\nonumber\\
 &-\frac{\Omega}{2}\left[e^{i\bar\gamma}\delta_{j-1,j^\prime} + \textrm{h.c.}\right]\label{EqnDisplaced}
\end{align}
each labeled by a quasi-momentum ${\hbar \bf q}$.  The first (kinetic energy) term in Eq.~(\ref{EqnDisplaced}) describes the 2D displaced parabolae depicted in Fig.~\ref{fig_setup}c.  In analogy to band-structure, the last (coupling) term in Eq.~(\ref{EqnDisplaced}) has the form of an $N$ site 1D periodic tight binding Hamiltonian with a ``magnetic flux'' $N\bar\gamma$ and a ``hopping'' matrix element $\Omega/2$, where internal atomic states play the role of lattice sites.  We diagonalize the coupling term, (a zero-order approximation suitable when $\Omega$ is much larger than all other parameters), by transforming into the basis conjugate to the spin-index $j$ with field operators
\begin{align*}
\hat{\varphi}^\dagger_\ell(\bf q) &= \frac{1}{N^{1/2}}\sum_{j=1}^N e^{i 2\pi\ell j/N} \hat{\bar\varphi}^\dagger_j(\bf q).
\end{align*}
The diagonalization provides the eigenenergies $E_\ell=-\Omega\cos(2\pi\ell/N-\bar\gamma)$ of the coupling Hamiltonian, where $\ell\in\{0,\cdots,N-1\}$ is analogous to the usual crystal momentum.  The ground state can be made two-fold degenerate by tuning $\bar\gamma$ to ``magic'' phases $\bar\gamma = 2\pi (p+1/2)/N$ for $p\in \mathbb{Z}$.  The manufactured degeneracy between states at $\ell=0$ and $1$ for $\bar\gamma=\pi/N$ is illustrated by Fig.~\ref{fig_energies} for $N=3$ and $4$.

When the displacement vectors ${\bf K}_j$ reside on the vertices of a regular polygon, ${\bf K}_j=-\kl\sin\left(2\pi j/N\right){\bf e}_x +\kl\cos\left(2\pi j/N\right){\bf e}_y$, the full Hamiltonian matrix is
\begin{align}
H_{\ell,\ell^\prime}({\bf q}) =&\left({\bf q}^2+1+E_\ell\right)\delta_{\ell,\ell^\prime}\nonumber\\
&+\left[\left(i q_x+q_y\right)\delta_{\ell-1,\ell^\prime}+\textrm{h.c.}\right] + \frac{\epsilon}{2}\delta_{\ell-N/2,\ell^\prime},
\label{DressedFullHamiltonian}
\end{align}
where momenta and energies are expressed in recoil units, $\kl$ and $\El=\hbar^2\kl^2/2m$ respectively.  Assuming $\bar\gamma\approx\pi/N$, we focus on the manifold of two nearly degenerate states with $\ell=0$ and $\ell=1$, yielding the pseudospins $\ket{\downarrow}$ and $\ket{\uparrow}$ depicted in Fig.~\ref{fig_energies}.

In what follows, we derive an effective $2\times2$ Hamiltonian $\check H\approx \check H^{(0)} + \check H^{(2)}+  \check H^{(3)}$ for this manifold up to third order in powers of $\Omega^{-1}$.  In the subspace spanned by the lowest energy pseudospin-pair we obtain (up to a constant) a zero-order Hamiltonian of the Rashba form
\begin{align}
\check H^{(0)}&=\left|{\bf q}\right|^2\check1+\left(\check\sigma_x q_y -\check\sigma_y q_x\right)+\frac{\Delta_{\rm Z}}{2}\check\sigma_z\label{EqnRashba}
\end{align}
with a Zeeman field $\Delta_{\rm Z}=E_1-E_0\approx-2\Omega\bar\gamma'\sin\left(\pi/N\right)$ generated by slight changes $\bar\gamma'=\bar\gamma-\pi/N$ from the magic phase.  At finite coupling $\Omega$, we adiabatically eliminate the excited states order-by-order in perturbation theory giving effective terms $H^{(n)}$ in the ground manifold Hamiltonian.

Since Eq.~(\ref{DressedFullHamiltonian}) is cyclic for $\epsilon=0$, we expect an energy shift at order $n=2$ in perturbation theory (effectively a Stark shift), and pseudospin-changing terms at order $n=N-1$.  These terms serve to restore the $N$-fold rotational symmetry absent from Eq.~(\ref{EqnRashba}) and in the analogous expressions of earlier proposals~\cite{Ruseckas2005,Stanescu2008,Juzeliunas2010}.  To understand the departure from the Rashba Hamiltonian, we will first consider the simpler $\epsilon=0$ case.

\begin{figure}[tb]
\begin{center}
\includegraphics[width=3.0in]{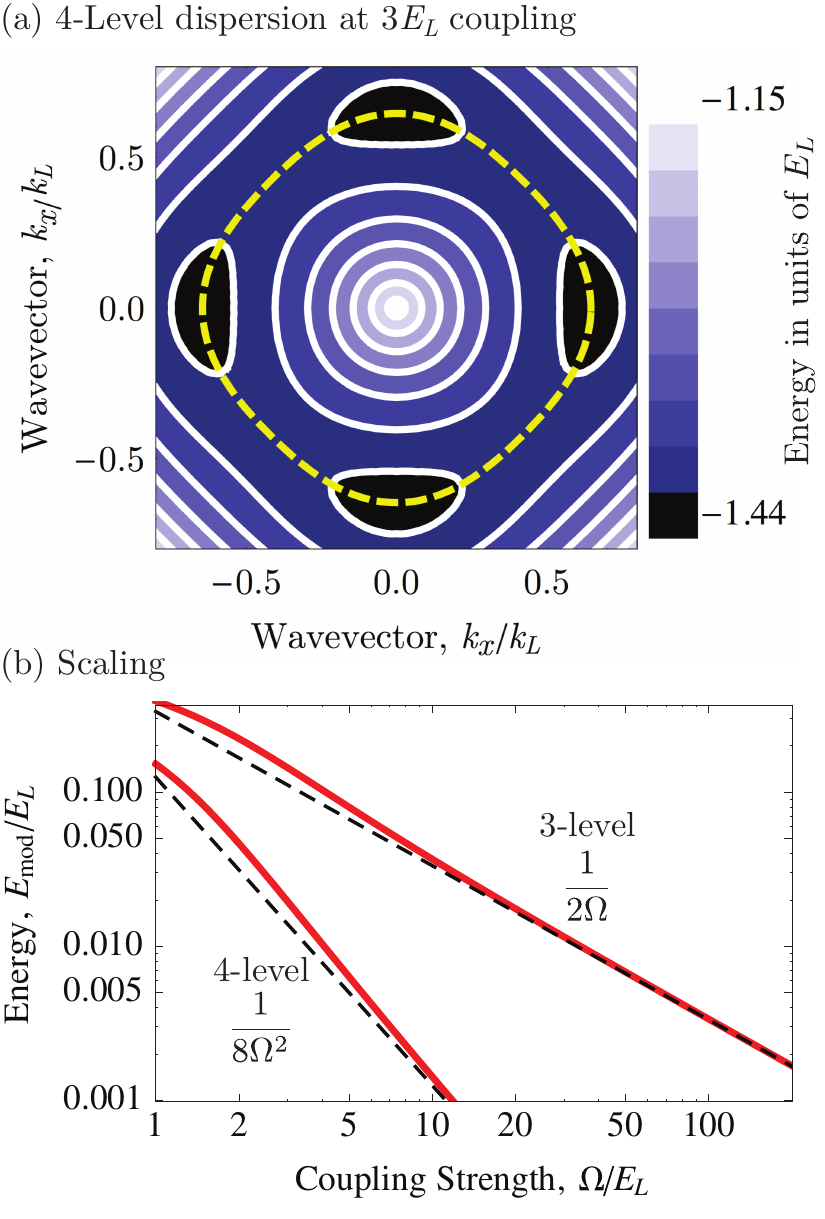}
\end{center}
\caption{a) Energy of the lowest eigenstate for $N=4$, $\Omega=3\El$, $\epsilon=0$, and $\bar\gamma=\pi/4$ plotted in the $q_x$-$q_y$ plane showing the four-fold rotational symmetry.  The graph is colored according to the energy, and white curves mark contours of equal energy.  The yellow dashed line depicts azimuthal modulation of radius in momentum of the energy-minimum.  b) Peak-to-peak magnitude of the azimuthal energy modulations $E_{\rm mod}$ plotted as a function of coupling $\Omega$, showing a rapid suppression for $N=4$ as compared to $N=3$.  The solid red curves, the result of exactly solving the full Hamiltonian, quickly converge to the perturbation result (black dashed lines).}
\label{fig_symmetry}
\end{figure}

For the $N=3$ case the second order effective Hamiltonian 
\begin{align*}
\check H^{(2)} &= -\frac{2}{3\Omega}\left[\left|{\bf q}\right|^2\check 1 + 2\check\sigma_x (q_y^2-q_x^2) + \check\sigma_y q_x q_y\right]
\end{align*}
restores the expected 3-fold symmetry.  For $N=4$, the second order term acts as a state-independent Stark shift, but the third order correction
\begin{align*}
\check H^{(3)} &= \frac{1}{2\Omega^2}\left[\check\sigma_x(q_y^3-3q_y q_x^2) - \check\sigma_y(q_x^3-3q^2_y q_x)\right]
\end{align*}
restores the 4-fold rotational symmetry.  This term is reminiscent of the cubic Dresselhaus SO coupling present in GaAs 2D electron systems~\cite{Stanescu2007a,Koralek2009}.  The Rashba Hamiltonian's [Eq.~(\ref{EqnRashba})] ground state energy is minimized on the ring where $|{\bf q}|=1/2$; the perturbative terms modulate both the momenta and energy where the minima occur~\footnote{In practice, the confining harmonic potential also lifts the Rashba Hamiltonian's ground state degeneracy at the level of the typical $\approx10\Hz$ trap frequencies~\cite{Larson2009}.  Thus we consider perturbative corrections below $0.001\El\approx h\times2.5\Hz$ as negligible.}.  Figure~\ref{fig_symmetry}a shows this modulated energy for the 4-level case at $\Omega=3\El$; Fig.~\ref{fig_symmetry}b plots the peak-to-peak amplitude of the energy modulations.  The perturbative analysis given by the dashed lines rapidly converges to the numerical solution to the full Hamiltonian given by the solid curves.  Since the magnitude of the modulations scale like $\Omega^{2-N}$, the 3-level case requires an impractically large $\Omega>100\El$ to reduce the corrections to the Rashba Hamiltonian below $10^{-3}\El$, while the 4-level case requires just $\Omega=10\El$.  Interestingly, the familiar tripod scheme~\cite{Ruseckas2005,Stanescu2008,Liu2009,Juzeliunas2010,Dalibard2010} reduces to our $N=3$ ring model when far detuned from the excited state, with $\bar\gamma=0$ for red detuning and $\bar\gamma=\pi/3$ for blue detuning.   By contrast, the corrections to the Rashba model converge as $\Omega^{-2}$ in the standard resonant tripod (a four level system).

The alternating detuning $\epsilon$ featured in Eq.~(\ref{EqnDisplaced}) leads to a super-lattice in the above mentioned band-structure analogy, and its contribution can be included exactly.  However, for a more painless description, we take $\epsilon/\Omega\ll1$, which for $N=4$ adds a tunable Dresselhaus term $\check H_D = \left(\El\epsilon/\sqrt{2}\Omega\kl\right)\left(\check\sigma_x q_y + \check\sigma_y q_x\right)$ at second order in $\Omega^{-1}$ (in original units).  Thus, our scheme produces both Rashba and Dresselhaus couplings with strengths $\alpha=\El/\kl$ and $\beta=\El\epsilon/\sqrt{2}\Omega\kl$ along with a $\hat z$-aligned Zeeman field $\Delta_{\rm Z}\approx-\sqrt{2}\Omega\bar\gamma^\prime$.  The laser configuration specifies $\alpha$; the alternating laser detunings set $\beta$; and $\Delta_{\rm Z}$ is controlled by the average phase $\bar\gamma$.  Together this allows for nearly complete control of the SO Hamiltonian.

\begin{figure}[tb]
\begin{center}
\includegraphics[width=3.3in]{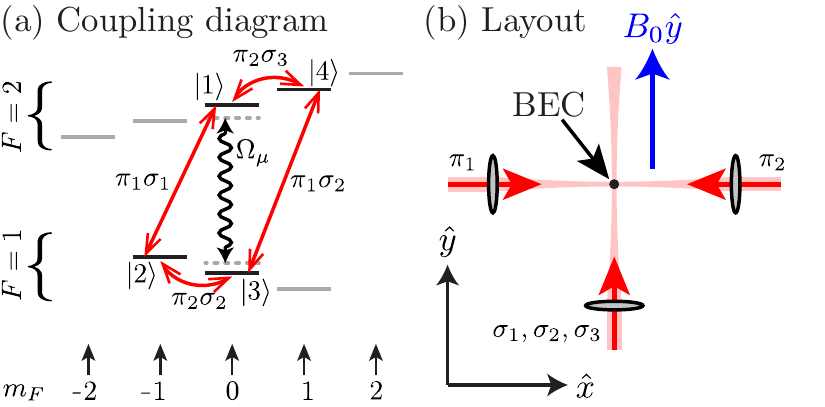}
\end{center}
\caption{\label{Rashba_geometry} 4-level ring coupling scheme in $\Rb87$ involving hyperfine states $\ket{F,m_F}$ Raman-coupled by a total of five lasers marked $\sigma_1$, $\sigma_2$, $\sigma_3$, $\pi_1$ and $\pi_2$.  (a) Level diagram:  Each red curve represents a two-photon Raman transition with polarizations as marked by the field-symbols.  The ring states are mapped to physical states according to $\ket{1}=\ket{2,0}$, $\ket{2}=\ket{1,-1}$, $\ket{3}=\ket{1,0}$, and $\ket{4}=\ket{2,1}$.  A $~6.8\GHz$ microwave field with coupling strength $\Omega_\mu$ ac Zeeman shifts the internal states $\ket{1,0}$ and $\ket{2,0}$.  (b) Schematic layout: a bias magnetic field $B_0\approx 0.2\mT$ lies along $\hat y$.  Lasers $\pi_{1}$ and $\pi_{2}$ are $\pi$ polarized.  Lasers $\sigma_{1}$, $\sigma_2$ and $\sigma_{3}$ are left circularly polarized and their relative phases define $N\bar\gamma=2\phi_{\sigma_{2}}-\phi_{\sigma_{1}}-\phi_{\sigma_{3}}$.}
\end{figure}

{\it Implementation}
In the following, we specialize to a 4-state configuration within the $F=1$ and $F=2$ hyperfine manifolds of $\Rb87$'s $^{5}\rm{S}_{1/2}$ ground electronic state (illustrated in Fig~\ref{Rashba_geometry}a).  Each pair of states is Raman coupled by two lasers tuned between the $^{5}\rm{P}_{3/2}$ and $^{5}\rm{P}_{1/2}$ atomic excited manifolds ($\El\approx h\times2500\Hz$).  Although there are nominally eight transitions coupling four states we reuse three coupling lasers.  The symbols $\sigma_1$, $\sigma_2$, $\sigma_3$, $\pi_1$ and $\pi_2$ each implicitly denote a laser frequency (that may be inferred from the level diagram in Fig~\ref{Rashba_geometry}a, the energy of each state $\ket{j}$ and the choice of 2-photon detuning for one set of Raman lasers) and a direction given by Fig~\ref{Rashba_geometry}b.  The first three symbols denote left circular polarization and their relative phases define $N\bar\gamma=2\phi_{\sigma_{2}}-\phi_{\sigma_{1}}-\phi_{\sigma_{3}}$.  Lasers  $\pi_1$ and $\pi_2$ are linearly polarized along the magnetic field and produce canceling phases in the Hamiltonian and thus do not contribute to the phase-sum.  Additionally, state labels will match those of the paper given ordering of frequency $\nu_{\pi_1}>\nu_{\sigma_3}>\nu_{\pi_2}>\nu_{\sigma_2}>\nu_{\sigma_1}$.  A bias field along $\hat y$ and an ac Zeeman shift provided by a microwave field along $\hat x$ produce a composite Zeeman shift that spectroscopically isolate~\cite{Lin2011} the hyperfine changing transitions by at least $10\Omega$.  In addition, $\ket{2}$ and $\ket{4}$ shift together with applied magnetic field, allowing straightforward control of the detuning $\epsilon$, thus contributing Dresselhaus coupling.




In practice, electronic equipment will introduce unwanted time-varying terms into the Hamiltonian. However, the effects of magnetic field detuning $\Delta\epsilon$, laser intensity shifts $\Delta\Omega$ and phase shifts $\Delta\bar\gamma$ can have negligible contributions in typical laboratory settings.  In the scheme shown in Fig.~\ref{Rashba_geometry}, time varying magnetic fields induce identical linear Zeeman shifts of $\ket{2}$ and $\ket{4}$ that cancel at lowest order in perturbation theory (while $\ket{1}$ and $\ket{3}$ are first order insensitive to magnetic fields).  These Zeeman shifts give an additional Dresselhaus contribution $\check H_{\Delta D} \propto \Delta\epsilon\left(\check\sigma_x q_y + \check\sigma_y q_x\right)$; near the minima of the Rashba Hamiltonian, the energy shift from the resulting Dresselhaus term would be $~0.0005\El$ for our $\Delta\epsilon\approx h\times50\Hz$ detuning noise amplitude~\cite{Lin2011}.


Shifts (resulting from laser intensity noise) in a single $\Omega_{j,j+1}$ matrix element add coupling terms $\check H_{\Delta\Omega}\approx-\sqrt{2}\Delta\Omega_{j,j+1} (\boldsymbol{K_{j}}+\boldsymbol{K_{j+1}})\cdot\check{\boldsymbol{\sigma}}/8 \kl$, where $\check{\boldsymbol{\sigma}}$ is the vector of Pauli matrices.  For lasers stabilized at the $0.1\%$ level, and with $\Omega=10\El$, the unwanted coupling terms have magnitude $~0.002\El$.


Phase shifts $\Delta\bar\gamma$ directly modulate the effective Zeeman fields $\check H_{\Delta\phi}\approx -\sqrt{2}\Omega \Delta\bar\gamma \check \sigma_z/2$ which opens a gap at the Dirac point.  For $\Rb87$, we require the resulting gap energy to be less than a typical $T=10\nK$ temperature.   For $\Omega=10\El$, the RMS phase noise must have an amplitude $\Delta\bar\gamma<0.04\ {\rm rad}$.  Even for independent lasers, this level of phase stability is routine~\cite{Cacciapuoti2005}.

{\it Discussion} Inspection of Eq.~(\ref{DressedFullHamiltonian}) shows that Dirac points are present for all $\Omega$ and $\epsilon$ provided $\bar\gamma=\pi/N$, even when perturbative corrections are important.  As a result, properties of fermion systems which depend only on the topology of the dressed-state dispersion may be insensitive to small corrections to Eq.~(\ref{DressedFullHamiltonian}).  In contrast, bosons generally condense at the energy minima.   Thus, for insufficient laser coupling, local minima may spoil correlation physics potentially arising from the Rashba Hamiltonian's degenerate ring of minima~\cite{Stanescu2008,Larson2009}.  

The proposed coupling scheme provides a robust platform for generating SO coupling for neutral atoms.  Because the two spin states are the lowest energy dressed states, atom-atom interactions cannot induce collisional decay~\cite{Spielman2006,Williams2011}.  In addition, this  technique can require considerably less laser intensity than prior far-detuned schemes to reach nearly pure SO coupled Hamiltonians, greatly reducing spontaneous emission.  Lastly our specific implementation uses only the $\delta m_F=0,\pm1$ Raman transitions allowed at large atomic detuning in the alkali atoms~\cite{Juzeliunas2010,Dalibard2010}.  


{\it Acknowledgments}
We thank KITP Santa Barbara (NSF Grant No. PHY05-51164) where this collaboration was initiated. D.L.C. and I.B.S. acknowledge the financial support of the NSF through the PFC at JQI, and the ARO with funds from both the Atomtronics MURI and the DARPA OLE Program.  G.J acknowledges the support from the EU project STREP NAMEQUAM.

\bibliography{RingCoupling}

\end{document}